\documentstyle[12pt]{article}

\newtheorem{theorem}{Theorem}
\newtheorem{lemma}{Lemma}

\begin{document}

\newcommand{\lap}{\bigtriangleup}
\def\be{\begin{equation}}
\def\ee{\end{equation}}
\def\bea{\begin{eqnarray}}
\def\eea{\end{eqnarray}}
\def\beas{\begin{eqnarray*}}
\def\eeas{\end{eqnarray*}}
\def\n#1{\vert #1 \vert}
\def\nn#1{{\Vert #1 \Vert}}
\def\R{{\rm I\kern-.1567em R}}
\def\N{{\rm I\kern-.1567em N}}
 
\def\supp{\mbox{\rm supp}\,}
\def\dist{\mbox{\rm dist}\,}

\def\suppi{\mbox{\scriptsize supp}\,}
\def\ekin{E_{\rm kin}}
\def\epot{E_{\rm pot}}

\def\D{{\cal D}}
\def\C{{\cal C}}
\def\X{{\cal X}}
\def\F{{\cal F}}
\def\P{{\cal P}}
\def\M{{\cal M}}
\def\Hc{{\cal H_C}}
\def\prfe{\hspace*{\fill} $\Box$

\smallskip \noindent}

\title{Stability of spherically symmetric
       steady states in galactic dynamics against general perturbations}
\author{ Gerhard Rein\\
         Mathematisches Institut 
         der Universit\"at M\"unchen\\
         Theresienstr. 39\\
         80333 M\"unchen, Germany}
\date{}
\maketitle

\begin{abstract}
Certain steady states of the Vlasov-Poisson system can be
characterized as minimizers of an energy-Casimir
functional, and this fact
implies a nonlinear stability property of such steady states. In previous investigations by {\sc Y.~Guo} and the author 
stability was obtained only with respect to spherically symmetric 
perturbations. In the present investigation we show how to remove this unphysical restriction.

\end{abstract}

\section{Introduction}
\setcounter{equation}{0}

In astrophysics the Vlasov-Poisson system 
\be \label{vlasov}
\partial_t f + v \cdot \partial_x f - \partial_x U \cdot 
\partial_v f = 0 ,
\ee
\be \label{poisson} 
\lap U = 4 \pi\, \rho,\ \lim_{\n{x} \to \infty} U(t,x) = 0 , 
\ee
\be \label{rho}
\rho(t,x)= \int f(t,x,v)dv ,
\ee
is used to model the time evolution of large stellar systems such as 
galaxies or a globular clusters.
Here $f = f(t,x,v)\geq 0$ denotes the density of the stars 
in phase space, $t \in \R$ denotes time, $x, v \in \R^3$ denote 
position and velocity respectively, $\rho$ is the spatial mass 
density, and $U$ the gravitational potential. 
The model neglects relativistic effects or collisions among the stars.

In a series of papers {\sc Y.~Guo} and the author have developed a variational
technique to construct stable steady states of this system,
cf.\ \cite{G1,GR1,GR2,R3}: Under some
assumptions on the function $Q$ the energy-Casimir functional
\be \label{encas}
\Hc (f):= \frac{1}{2} \int\!\!\int \n{v}^2 f(x,v)\,dv\,dx
- \frac{1}{8\pi} \int |\nabla U_f (x)|^2 dx
+ \int\!\!\int Q(f(x,v))\,dv\,dx
\ee
has a minimizer $f_0$ in some function set, this minimizer is easily seen to
be a steady state of the Vlasov-Poisson system, and its minimizing
property implies nonlinear stability. These investigations were restricted
to the case of spherical symmetry (or ``flat'' axial symmetry in the case
of \cite{R3}), and the aim of the present paper is to remove this restriction.
A physically realistic perturbation, say by the gravitational
pull of some distant galaxy, is hardly spherically symmetric.

The variational equation for the minimizer $f_0$ shows that
$f_0$ depends only on the particle energy    
\be \label{parten}
E=\frac{1}{2}|v|^2 + U_0 (x), 
\ee
which is a conserved quantity along characteristics;
$U_0$ is the potential induced by $f_0$. In the case of spherical symmetry
steady states may also depend on a further conserved quantity,
the modulus of angular momentum squared:
\be \label{angmom}
L:= \n{x}^2 \n{v}^2 - (x \cdot v)^2 .
\ee
To obtain such steady states, the function $Q$ in (\ref{encas})
must also depend on $L$, and \cite{G1,GR1,GR2} actually treated
this more general case. However, the present investigation does
not cover the case of $L$-dependent steady states, 
the reason being as follows: For the minimizing property of $f_0$ 
to imply stability $\Hc$ should be conserved along solutions
of the Vlasov-Poisson system. This is the case for not
necessarily symmetric solutions if $\Hc$ is as in (\ref{encas}),
but it requires spherical symmetry if $Q$ depends on $L$.

To put the present investigation into perspective we compare
the variational approach sketched above with other approaches. 
To this end we recall the well known class of polytropic steady
states where
\be \label{poly}
f_0 (x,v) = (E_0 - E)_+^k L^l.
\ee
Here $(\cdot)_+$ denotes the positive part, $E_0\in \R$ is a constant,
and  $k>-1,\ l>-1,\ k+l+1/2 >0,\ k< 3l +7/2$; only for this range of exponents
do these steady states have compact support and finite mass. 
The first nonlinear stability result for the Vlasov-Poisson
system in the present stellar dynamics case is due to {\sc G.~Wolansky}
\cite{Wo}. It is restricted to spherically symmetric perturbations
of the polytropes  
with exponents $l > -1$, $0< k < l+3/2$ with $k \neq - l - 1/2$
and uses a variational approach for a reduced functional which is
not defined on a set of phase space densities $f$ but on a set
of mass functions $M(r):=\int_{\n{y} \leq r} \rho(y)\,dy$, $r \geq 0$ 
the radial
coordinate. In particular, is does not yield a stability estimate
directly for the phase space distribution $f$.
In a recent paper {\sc Y.-H.~Wan} proves stability
by a careful investigation of the quadratic and and higher order parts 
in a Taylor expansion of $\Hc$ about a steady state. He has to assume the
existence of the steady state, requires a strong condition on $f_0$
which is satisfied by the polytropes only for $k=1$ and $l=0$,
but his arguments do not require spherical symmetry of the admissible perturbations, cf.\ \cite{Wa}.
Finally, the approach in \cite{G1,GR1,GR2} gives the existence of the steady
states (and actually provides new ones), covers the polytropes
for $l>-1$ and $0 < k < l + 3/2$, and, as we believe, 
has the simplest proof of the
three approaches. With the present investigation we remove the only restriction
this approach had so far when compared with \cite{Wa}, namely
spherical symmetry of the admissible perturbations.
We also mention \cite{A} where stability for the limiting case $k=7/2$ and
$l=0$ of polytropes which have finite
mass but infinite support is treated by a variational technique.

The paper proceeds as follows: In the next section we establish 
some preliminary estimates which in particular 
show that $\Hc$ is bounded from below and 
the positive terms in $\Hc$ are bounded along minimizing sequences.
In Section~3 the existence of a minimizer of $\Hc$ is established.
Most of the technical steps can be taken over
from \cite{GR1,GR2}, since in these papers spherical symmetry  was only used
to prevent mass from running off to spatial infinity along a minimizing
sequence. To control this in the nonsymmetric case we use a 
concentration-compactness lemma due to {\sc P.-L.~Lions}.
In Section~4 we show that such minimizers are spherically 
symmetric steady states of the Vlasov-Poisson
system with finite mass and compact support.
The stability properties of the steady states are then discussed in the last section. Here we point out one problem: 
If $f_0$ is a steady state then 
$f_0(x+V\,t, v+V)$ for any given velocity $V \in \R^3$ is a
solution of the Vlasov-Poisson system which for $V$ small starts
close to $f_0$, but travels away from $f_0$ at a linear
rate in $t$. This trivial ``instability'', which cannot be present
for spherically symmetric perturbations, has to be dealt with,
and incidentally, both \cite{Wa} and the present paper handle this
by comparing $f_0$ with an appropriate shift in $x$-space
of the time dependent perturbed solution $f(t)$.
In our case this shift arises from the application of the
concentration-compactness lemma.

We conclude the introduction with some further references.
Global classical solutions 
to the initial value problem for the Vlasov-Poisson system were first 
established in \cite{P}, cf.\ also \cite{S}.  
Many references 
to discussions of the stability problem in the astrophysics
literature can be found in the monograph \cite{FP}. 
A rigorous  investigation of linearized stability 
is given in \cite{BMR}. For the plasma physics case,
where the sign in the Poisson equation (\ref{poisson}) is
reversed, the stability problem is much easier and better
understood. We refer to \cite{BRV,GS1,GS2,R1}.  
Finally, a very general condition
which guarantees finite mass and compact support of steady
states, but not their stability, is established in \cite{RR}.
   
\section{Preliminaries}
\setcounter{equation}{0}

For a measurable function $f=f(x,v)$ we define
\[
\rho_f (x):= \int f(x,v)\, dv,\ x \in \R^3,
\]
and
\[  
U_f :=  - \rho_f \ast \frac{1}{\n{\cdot}}.
\]
As to the existence of this convolution see Lemma~\ref{rhoest} below.
Next we define
\beas
\ekin (f)
&:=&
\frac{1}{2} \int\!\!\int \n{v}^2 f(x,v)\,dv\,dx,\\
\epot (f)
&:=&
- \frac{1}{8\pi} \int |\nabla U_f (x)|^2 dx = 
- \frac{1}{2} \int\!\!\int\frac{\rho_f(x) \rho_f(y)}{\n{x-y}}dx\,dy ,\\
\C(f)
&:=&
\int\!\!\int Q(f(x,v))\,dv\,dx,
\eeas
and
\[
\Hc(f) :=
\C(f) + \ekin(f)  + \epot (f),\ \P(f) :=
\C(f) + \ekin(f),
\]
where 
$Q$ is a given function satisfying certain assumptions specified 
below.
Note that $\P$ is the positive part of the energy-Casimir functional $\Hc$.  
We will minimize $\Hc$ over the set
\be \label{spacedef}
\F_M := \Bigl\{ f \in L^1(\R^6) 
\mid
f \geq 0,\ \int\!\!\int f dv\,dx = M,\ \P(f) < \infty\Bigr\},
\ee
where $M>0$ is prescribed. The function $Q$ which determines the Casimir functional has to satisfy
the following  

\smallskip
\noindent {\bf Assumptions on $Q$}: 
$Q \in C^1 ([0,\infty[)\cap 
C^2 (]0,\infty[)$, $Q \geq 0$, and there exist
constants $C_1,\,C_2 >0$, $F_0 >0$, and 
$0 < k_1,\,k_2,\,k_3 < 3/2$ such that:
\begin{itemize}
\item[(Q1)]
$Q(f) \geq C_1 f^{1+1/{k_1}},\ f \geq F_0$.
\item[(Q2)]
$Q(f) \leq C_2 f^{1+1/{k_2}},\ 0 \leq f \leq F_0$.
\item[(Q3)]
$Q(\lambda f) \geq \lambda^{1+1/{k_3}} Q(f),\ f \geq 0,\ 
0 \leq \lambda \leq 1$. 
\item[(Q4)]
$ Q''(f) > 0,\ f > 0$, and $ Q'(0) = 0$.
\end{itemize}
On their support
the steady states obtained later will be of the form 
\[
f_0 (x,v) = (Q')^{-1}(E_0 -E) 
\]
with some $E_0<0$ and $E$ as defined in 
(\ref{parten}); under the assumptions above 
$Q'$ is strictly increasing with range $[0,\infty[$.
A typical example of a function $Q$ satisfying the assumptions is
\be \label{qpoly}
Q(f) = f^{1+1/{k}},\ f \geq 0,
\ee
with $0 < k < 3/2$ which leads to a steady state of polytropic form (\ref{poly}). 

We collect some estimates
for $\rho_f$ and $U_f$ induced by an element $f \in \F_M$.
As in the rest of the paper constants
denoted by $C$ are positive, may depend on $M$ and $Q$, and may 
change their value from line to line.

\begin{lemma} \label{rhoest}
Let $n_1 := k_1 +3/2$ so that $1+1/n_1 > 4/3$. 
Then for any $f \in \F_M$ the following holds:
\begin{itemize}
\item[{\rm (a)}] 
$f \in L^{1+1/k_1} (\R^6)$ with
\[
\int\!\!\int f^{1+1/k_1} dv\,dx \leq C (1 + \P(f)).
\]
\item[{\rm (b)}]
$\rho_f \in L^{1+1/n_1} (\R^3)$ with
\[
\int \rho_f^{1+1/n_1} dx \le C 
\left( 1 + \P (f) \right).
\]
\item[{\rm (c)}]
$U_f \in L^{12}(\R^3)$ with $\nabla U_f \in L^{2}(\R^3)$, and
\[
\int \n{\nabla U_f}^2 dx \leq C \nn{\rho_f}_{6/5}^2
\leq C \nn{\rho_f}_{1+1/n_1}^{(n_1+1)/3}.
\]
The two representations of $\epot (f)$ stated above are indeed equal.
\end{itemize}
\end{lemma}
\noindent
{\bf Proof.} As to (a) and (b)
we refer to \cite[Lemma~1]{GR1} or \cite[Lemma~1]{GR2}.
The estimates for $U_f$ follow from the generalized Young's inequality,
and the equality of the two representations for $\epot (f)$ follows
by regularizing $\rho_f$, integrating by parts, and then passing to the limit.
\prfe

An an immediate corollary of the lemma above one can show that
on $\F_M$ the functional $\Hc$ is bounded from below in such a way
that $\P$---and thus certain norms of $f$ and $\rho_f$---remain bounded 
along minimizing sequences---note that $n_1 < 3$:

\begin{lemma} \label{lower}
For every $M>0$ there exists a constant $C>0$ such that
\[
\Hc (f) \geq  \P(f) - C \left( 1 + \P (f) \right)^{n_1/3},\ f \in \F_M,
\]
in particular,
\[
h_M := \inf_{\F_M} \Hc  > - \infty .
\]
\end{lemma}
 
\section{Existence of minimizers }
\setcounter{equation}{0}

The behaviour of $\Hc$ and $M$ under scaling transformations can be
used to show that $h_M$ is negative and to
relate the $h_M$'s for different values of $M$:

\begin{lemma} \label{scaling}
\begin{itemize}
\item[{\rm (a)}]
Let $M>0$. Then $-\infty < h_M  < 0$.
\item[{\rm (b)}]
There exists $\alpha >0$ such that for all $ 0< M_1 \leq M_2$,
\[
h_{M_1}\ge \left( {{M_1}\over{M_2}}\right )^{1+\alpha}h_{M_2}.
\]
\end{itemize}
\end{lemma}

For the proof we refer to \cite[Lemma~4]{GR1} or \cite[Lemma~4]{GR2};
spherical symmetry was not used in those proofs,
cf.\ also \cite[Lemma~4]{R3} where we explicitly kept track of where
symmetry was used. A simple consequence of part (b) of the lemma above
is that
\[
h_{M} < h_{M-m} + h_m,\ 0 < m < M,
\]
which is condition (S.2) in \cite[Theorem II.1]{L}, but we prefer to work 
with (b). 

For easier reference 
we state the concentration-compactness lemma
which replaces the splitting estimates used in
\cite{GR1,GR2}. 
This is Lemma I.~1 in \cite{L}, cf.\ also
\cite[4.3]{St}. The fact that we have functions
of two variables $x$ and $v$ but consider the various balls 
\[
B_R:=\{ x \in \R^3 \mid \n{x} \leq R \}
\]
only in $x$-space requires no changes in the proof.

\begin{lemma} \label{cc}
Let $(f_n) \subset L^1(\R^6)$ with $f_n \geq 0$ and $\int f_n =M$,
$n \in \N$.
Then there exists a subsequence $(f_{n_k})$ such that one of the following assertions holds:
\begin{itemize}
\item[{\rm (i)}]
$\exists (a_k) \subset \R^3\ \forall \epsilon >0 \ \exists
R>0,\ k_0 \in \N$:
\[
\int_{a_k + B_R} \int f_{n_k} dv\,dx \geq M - \epsilon,\ k \geq k_0.
\]
\item[{\rm (ii)}]
$\forall R>0$:
\[\lim_{k\to \infty} \sup_{a \in \R^3} \int_{a+B_R} \int f_{n_k} dv\,dx = 0.
\]
\item[{\rm (iii)}]
$\exists m \in ]0,M[ \ \forall \epsilon > 0\ \exists k_0 \in \N,\
(f_k^1),\ (f_k^2) \subset L^1(\R^6)$:
\[
\nn{f_{n_k} - (f_k^1 + f_k^2)}_1 \leq \epsilon,\
\left| \int f_k^1 -m \right| \leq \epsilon,\
\left| \int f_k^2 -(M-m) \right| \leq \epsilon,\ k \geq k_0,
\]
\[
\dist\left(\supp f_k^1,\supp f_k^2\right) \to \infty,\ k \to \infty,
\]
and
\[
0 \leq f_k^1,\, f_k^2,\, g_k \leq f_{n_k},\ f_k^1\, f_k^2 = 
f_k^1\, g_k = f_k^2\, g_k = 0\ \mbox{a.~e.},\ k \geq k_0
\]
where $g_k := f_{n_k} - f_k^1 - f_k^2$.
\end{itemize}
\end{lemma}
In the case of a minimizing sequence for
$\Hc$ Lemma~\ref{scaling} can be used to exclude possibilities
(ii) and (iii) in the previous lemma:

\begin{lemma} \label{concentrate}
Let $(f_n) \subset \F_M$ be a minimizing sequence of $\Hc$.
Then in Lemma~\ref{cc} only (i) holds.
\end{lemma}

\noindent
{\bf Proof.} 
For $R>1$ define
\[
K_R(x):= \left\{ \begin{array}{ccl}
1/\n{x}&,& 1/R \leq \n{x} \leq R,\\
R &,& \n{x} < 1/R,\\
0 &,& \n{x} > R ,
\end{array} \right.
\]
and
\[
F_R (x) := \frac{1}{\n{x}} 1_{\{\n{x} > R\}} (x),\ 
G_R (x) := \left(\frac{1}{\n{x}} - R \right) 1_{\{\n{x} < 1/R\}} (x)
\]
so that
\be \label{splitgreen}
\frac{1}{\n{x}} = K_R (x) + F_R (x) + G_R (x),\ x \in \R^3 .
\ee
Here $1_A$ denotes the indicator function of the set $A$.
Assume (ii) holds and split
\[
\frac{1}{4 \pi} \int \n{\nabla U_n}^2 dx 
= \int\!\!\int \frac{\rho_n (x) \rho_n (y)}{\n{x-y}} dy\,dx = I_1 + I_2 + I_3
\]
according to (\ref{splitgreen}). 
Since $(\rho_n)$ is bounded in $L^{4/3}(\R^3)$ and $\nn{\rho_n}_1 = M,\
n \in \N$, we find
\beas
\n{I_1} 
&\leq&
R \int\!\!\int_{\n{x-y} < R} \rho_n(x)\,\rho_n(y)\,dx\,dy \leq R M \sup_{y \in \R^3} \int_{y+B_r} \rho_n (x)\, dx,\\
\n{I_2}
&\leq&
\frac{1}{R} \int\!\!\int \rho_n(x)\,\rho_n(y)\,dx\,dy = M^2 R^{-1},\\
\n{I_3}
&\leq&
\nn{\rho_n}_{4/3} \nn{\rho_n \ast G_R}_4 \leq C \nn{\rho_n}_{4/3}^2 \nn{G_R}_2 \leq
C R^{-1/2};
\eeas
for the last term we used H\"older's and Young's inequality.
Since this holds for any $R>1$, we conclude by (ii) that
$\epot(f_{n_k}) \to 0$ along the subsequence obtained in Lemma~\ref{cc}.
Hence $h_M \geq 0$, a contradiction to Lemma~\ref{scaling} (a).

Assume that (iii) holds. We denote the subsequence
obtained in Lemma~\ref{cc}
by $(f_n)$. Let $m \in ]0,M[$ be according to (iii) and $\epsilon >0$
arbitrary. With 
$m_n := \int \rho_n^1,\ M_n := \int \rho_n^2$,
obvious definitions for $\rho_n^i$ and $\sigma_n := \int g_n dv$
we have
\[
\n{m_n - m} \leq \epsilon,\ \n{M_n - (M-m)} \leq \epsilon
\]
and
\[
\P(f_n) = \P(f_n^1 + f_n^2 + g_n) = \P(f_n^1) + \P(f_n^2) + \P(g_n) 
\geq \P(f_n^1) + \P(f_n^2).
\]
Moreover
\[
\epot(\rho_n) = \epot(\rho_n^1) + \epot(\rho_n^2) -I_1 - I_2 + I_3
\]
where
\beas
I_1 
&:=& \int\!\!\int\frac{\rho_n^1(x)\rho_n^2(y)}{\n{x-y}}dx\,dy,\\
I_2 
&:=&
\int\!\!\int\frac{\rho_n (x)\sigma_n(y)}{\n{x-y}}dx\,dy,\\
I_3 
&:=& 
\frac{1}{2} \int\!\!\int\frac{\sigma_n(x)\sigma_n(y)}{\n{x-y}}dx\,dy .
\eeas
To estimate $I_1$ observe that for $n$ sufficiently large,
$\dist(\supp \rho_n^1,\supp \rho_n^2) > 1/\epsilon$ so that
\[
\n{I_1} \leq M^2 \epsilon.
\]
To estimate $I_2$ use the generalized Young's inequality and interpolation to
find
\[
\n{I_2} \leq C \nn{\rho_n}_{6/5} \nn{\sigma_n}_{6/5} \leq C \nn{\sigma_n}_1^{1/3}
\leq C \epsilon^{1/3}.
\]
As to $I_3$ it suffices to observe that this term is nonnegative.
Thus for any $\epsilon < 1$ and all sufficiently large $n$ we find,
using Lemma~\ref{scaling} (b),
\beas
h_M 
&\geq&
\P(f_n) + \epot(f_n) - \epsilon \\
&\geq&
\P(f_n^1) + \P(f_n^2) + \epot(f_n^1) + \epot(f_n^2) - C \epsilon^{1/3}\\
&\geq&
h_{m_n} + h_{M_n} - C \epsilon^{1/3}\\
&\geq&
\left[\left(\frac{m_n}{M}\right)^{1+\alpha} +
\left(\frac{M_n}{M}\right)^{1+\alpha} \right] h_M - C \epsilon^{1/3};
\eeas
clearly $0<m_n,\, M_n < M$ for $\epsilon >0$ sufficiently small.
To continue we define 
\[
C_\alpha := - \inf_{x \in ]0,1[}
\frac{(1-x)^{1+\alpha}+x^{1+\alpha}-1}{(1-x)x} > 0. 
\]
Since $h_M < 0$ it follows that
\beas
1 
&\leq&
\left(\frac{m_n}{M}\right)^{1+\alpha} +
\left(\frac{M_n}{M}\right)^{1+\alpha}  + C \epsilon^{1/3}\\
&=&
\left(\frac{m}{M}\right)^{1+\alpha} +
\left(\frac{M-m}{M}\right)^{1+\alpha}  + C \epsilon^{1/3}\\
&&
{}+ \left(\frac{m_n}{M}\right)^{1+\alpha} - \left(\frac{m}{M}\right)^{1+\alpha} 
 +
\left(\frac{M_n}{M}\right)^{1+\alpha} - \left(\frac{M-m}{M}\right)^{1+\alpha}\\
&\leq&
1 - C_\alpha \left(1-\frac{m}{M}\right) \frac{m}{M}  + C \epsilon^{1/3} 
+ 2 \frac{1+\alpha}{M} \epsilon,\ \epsilon \in ]0,1[,
\eeas
and this is a contradiction.
Thus only assertion (i) can hold. \prfe

\begin{theorem} \label{exminim}
Let $M>0$.
Let $(f_n) \subset \F_M$ be a minimizing sequence of 
$\Hc$. Then there is a minimizer $f_0\in \F_M$, a subsequence
$(f_{n_k})$, and a sequence $(a_k) \subset \R^3$ such that 
\[
\Hc (f_0) = \inf_{\F_M} \Hc =: h_M
\] 
and $f^{a_k}_{n_k} \rightharpoonup f_0$ weakly in 
$L^{1+1/k_1} (\R^6)$.
For the induced potentials we have
$\nabla U^{a_k}_{n_k} \to \nabla U_0$ strongly in $L^2 (\R^3)$.
Here $f^a(x,v) := f(x+a,v)$.
\end{theorem}

\noindent
{\bf Proof.} 
Let $(f_n)$ be a minimizing sequence.
Use Lemma~\ref{concentrate} to choose a
subsequence, denoted by $(f_n)$ again and a sequence
$(a_n) \subset \R^3$ such that (i) in Lemma~\ref{cc}
holds. Let $\bar f_n (x,v):= f_n(x+a_n,v)$. This is again
a minimizing sequence, because $\Hc$ is translation invariant.
By Lemma~\ref{lower}, $(\P(\bar f_n))$ is bounded  
and thus $(\bar f_n)$ is bounded in $L^{1+1/k_1} (\R^6)$. 
Thus there exists a weakly convergent
subsequence, denoted by $(\bar f_n)$ again:
\[
\bar f_n \rightharpoonup f_0\ \mbox{weakly in }\ L^{1+1/k_1} (\R^6).
\]
Clearly, $f_0 \geq 0$ a.~e.
By Lemma~\ref{rhoest} $(\bar \rho_n)=(\rho_{\bar f_n})$ is bounded in 
$L^{1+1/n_1} (\R^3)$.
After extracting a further subsequence
\[
\bar \rho_n \rightharpoonup \rho_0:=\rho_{f_0} \
\mbox{weakly in }\ L^{1+1/n_1} (\R^3) .
\]
Also by weak convergence 
$\ekin(f_0) \leq \liminf_{n \to \infty} \ekin(\bar f_n)$. 
By (Q4) the functional $\C$ is convex.
Thus by Mazur's Lemma and Fatou's Lemma  
\[
\C(f_0) \leq \limsup_{n \to \infty}\, \C(\bar f_n), 
\]
and 
\[
\P(f_0) \leq \limsup_{n \to \infty}\, \P(\bar f_n) .
\]
We show that $f_0 \in \F_M$. Let $\epsilon >0$. By (i) in Lemma~\ref{cc}
and the boundedness of $\ekin(\bar f_n)$ 
there exists $R>0$ such that
\[
M \geq
\int_{B_R} \int f_0 dv\,dx \geq M - \epsilon 
\]
which implies that $\int\int f_0 = M$ 
and $f_0 \in \F_M$. 
It remains to deal with the potential energy:
\beas
\frac{1}{8\pi} \int \left| \nabla U_{\bar f_n} - \nabla U_0 \right|^2 dx
&=&
\frac{1}{2} \int\!\!\int\frac{(\bar \rho_n (x) - \rho_0 (x))
(\bar \rho_n (y) - \rho_0 (y))}{\n{x-y}}dx\,dy\\
&=&
I_1 + I_2 + I_3
\eeas
where the latter integral is split according to (\ref{splitgreen}).

{\em Estimate of $I_1$}: Define
\[
U_n^R := - \bar \rho_n \ast K_R,\ U_0^R := - \rho_0 \ast K_R,\ R>0.
\]
Since $K_R \in L^1 (\R^3) \cap L^\infty (\R^3)$ weak convergence of 
$\bar \rho_n$ implies that for any $R>0$,
\[
U_n^R \to U_0^R,\ n \to \infty,\ \mbox{pointwise on}\ \R^3 .
\] 
Since $\int U_n^R = \int U_0^R$ we find
$U_n^R \to U_0^R$ in $L^1(\R^3)$ as $n \to \infty$ for any $R>0$.
Now H\"older's inequality, an interpolation argument, and Young's inequality
together with the boundedness of the $\rho$'s in $L^{4/3}(\R^3)$ imply that
\beas
\n{I_1}
&\leq&
\nn{\bar \rho_n - \rho_0}_{4/3} \nn{U_n^R - U_0^R}_{4} \leq
C \nn{U_n^R - U_0^R}_{12}^{9/11} \nn{U_n^R-U_0^R}_1^{2/11}\\
&\leq&
C \nn{\bar \rho_n - \rho_0}_{4/3}^{9/11} \nn{U_n^R - U_0^R}_{1}^{2/11}
\to
0,\ n \to \infty,\ R>0.
\eeas
Obviously
\[
\n{I_2} \leq \frac{4 M^2}{R},
\]
and again by H\"older's and Young's inequality
\[
\n{I_3} \leq \nn{\bar \rho_n - \rho_0}_{4/3} 
\nn{(\bar \rho_n - \rho_0)\ast G_R}_4
\leq
C \nn{G_R}_2 \leq C R^{-1/2}.
\]
Thus
$\nabla \bar U_{\bar f_n} \to \nabla U_0$ in $L^2(\R^3)$ for $n \to \infty$,
and the proof is complete. \prfe

\section{Properties of minimizers}
\setcounter{equation}{0}

\begin{theorem} \label{propminim}
Let $f_0 \in \F_M$ be a minimizer of $\Hc$. Then
\[
f_0 (x,v)=\left\{
\begin{array}{ccl} 
(Q')^{-1}(E_0-E) &,& E_0 - E > 0,\\
0 &,& E_0 - E \leq 0
\end{array}
\right.
\]
where
\[
E := \frac{1}{2} \n{v}^2 + U_0 (x),
\]
\[
E_0 := {1\over M} \int\!\!\int \left(Q'(f_0)
+ E \right)\,f_0\,dv\,dx ,
\]
and $U_0$ is the potential induced by $f_0$.
In particular, $f_0$ is a steady state of the Vlasov-Poisson
system.
\end{theorem}

For the proof we refer to \cite[Thm.~2]{GR1} where a somewhat stronger
condition (Q4) was used or \cite[Thm.~2]{GR2} where (Q4) is as in the present
paper;
if anything, the fact that we do not require spherical symmetry
of the functions in $\F_M$ makes the proof of this theorem easier.

There now arise a couple of questions which are all interrelated:
Firstly, in which sense does $f_0$ satisfy the stationary Vlasov-Poisson
system? Up to now, the Poisson equation holds in the sense of distributions,
and the Vlasov equation in the sense that $f_0$ is constant along characteristics, but $\nabla U_0$ is not sufficiently regular to define
classical characteristics to begin with. Secondly, we know
that if we minimize $\Hc$ over the space of spherically symmetric
functions in $\F_M$ we obtain a spherically symmetric minimizer
with compact support. 
Are the minimizers that we obtain in the present, more general context
still spherically symmetric and compactly supported? 
Are they unique? 
These questions are considered next; $C^k_c$ and $C^k_b$ denote the space
of $C^k$ functions with compact support and with bounded derivatives
up to order $k$, respectively: 

\begin{theorem} \label{minreg}
Let $f_0 \in \F_M$ be a minimizer of $\Hc$ so that
by Theorem~\ref{propminim} $f_0(x,v) = \phi (E)$
with $\phi$ determined by $Q$.
\begin{itemize}
\item[{\rm (a)}]
Assume that
\[
\phi(E) \leq C_1' (E_0-E)^{k_1},\ E \to -\infty
\]
and
\[
\phi(E) \geq C_2' (E_0-E)^{k_2},\ E \to E_0 -
\]
for positive constants $C_1',\ C_2'$
(as is illustrated by the polytropes
these assumptions are compatible with the general assumptions on $Q$). 
Then $E_0 < 0$, $\rho_0 \in C^1_c (\R^3)$, $U_0\in C^2_b (\R^3)$
with $\lim_{\n{x} \to \infty} U_0(x) =0$, and the steady state 
is spherically symmetric with respect to some point in $\R^3$.
\item[{\rm (b)}]
If in particular $Q(f) = f^{1+1/k},\ f \geq 0$,
with $0 < k < 3/2$ then up to a shift 
in $x$-space the minimizer is unique in $\F_M$.
\end{itemize}
\end{theorem}

\noindent
{\bf Proof.}
To prove part (a) the basic idea is to use Sobolev embedding 
to obtain the desired regularity,
establish the appropriate behaviour of $U_0$ at infinity, and then
apply a result by {\sc Gidas, Ni} and {\sc Nirenberg} to conclude the
spherical symmetry, cf.\ \cite[Thm.~4]{GNN}.
First we show that
\be \label{udecbelow}
-U_0 (x) \geq \frac{M}{3 \n{x}},\ \n{x} \to \infty .
\ee
To see this, choose $R>0$ such that
\[
\int_{\n{y} \leq R} \rho_0 (y)\, dy > \frac{M}{2}.
\]
Since
\[
\left| \frac{1}{\n{x-y}} - \frac{1}{\n{x}} \right| \leq \frac{R}{(\n{x}-R)^2},\
\n{x} \geq 2 R,\ \n{y} \leq R
\]
we obtain (\ref{udecbelow}) by restricting the
convolution integral defining $U_0$ to the ball $\{\n{y} \leq R\}$
and expanding the kernel as indicated.

Next we claim that 
\be \label{lpest}
\rho_0 \in L^4 (\R^3) .
\ee
To see this, we observe that $f_0$ depends 
only on the particle energy $E$ via the function $\phi$, and thus
\be \label{rhodar}
\rho_0 (x) = h_\phi (U_0 (x)),\ x \in \R^3
\ee
where
\be \label{hdef}
h_\phi (u) :=  
4 \pi \sqrt{2} \int_u^\infty \phi(E)\, \sqrt{E-u}\, dE,\
u \in \R;
\ee
note that $h_\phi (u) = 0$ for $u \geq E_0$.
The general assumptions on $Q$ and the additional assumption in the theorem
imply that
\[
h_\phi (u) \leq C \left( 1+(E_0 - u)^{k_1 +3/2}\right),\ u \leq E_0.
\]
If we use this estimate on the set where $\rho_0$ is large---this set 
has finite measure---and the integrability of $\rho_0$ on the complement
we find that
\[
\int \rho_0 (x)^p \, dx \leq C + \int (-U_0 (x))^{(k_1 + 3/2) p} dx,
\]
and since by Lemma~\ref{rhoest} (c)
$U_0 \in L^{12}(\R^3)$ this is finite for $p=12/(k_1 + 3/2) > 4$.

The next step is to show that
\be \label{potdec}
U_0 \in L^\infty (\R^3),\ U_0 (x) \to 0,\ \n{x} \to \infty.
\ee
To see this we split the potential in the following way:
\beas
- U_0 (x)
&=&
\int_{\n{x-y} < 1/R} \frac{\rho_0 (y)}{\n{x-y}}dy
+  \int_{1/R \leq \n{x-y} < R} \cdots +  \int_{\n{x-y} \geq R} \cdots\\
&\leq& 
C \nn{\rho_0}_4 \left( \int_0^{1/R} r^{2-4/3}dr\right)^{3/4}
+
R \int_{\n{y} \geq \n{x} - R} \rho_0(y)\, dy +\frac{M}{R},\ 
\n{x} \geq R,
\eeas
and since $R>1$ is arbitrary and $\rho_0 \in L^1(\R^3)\cap L^4(\R^3)$ assertion (\ref{potdec})
follows.

We are now in the position to show that 
\be \label{compsupp}
E_0 < 0\ \mbox{and}\ \supp \rho_0 \ \mbox{is compact} .
\ee
By (\ref{rhodar}) and (\ref{potdec}) the second assertion follows 
from the first. Assuming that $E_0>0$ immediately implies
that $\rho_0 (x) \geq h_\phi (E_0/2) >0$ for all sufficiently
large $\n{x}$ which contradicts the integrability of $\rho_0$.
Assume that $E_0=0$. Then the estimate for $\phi$ from below implies
\[
h_\phi (u) \geq C (-u)^{k_2 + 3/2},\ u \to 0-.
\]
But then (\ref{udecbelow}) implies that
\[
\rho_0 (x) \geq C \n{x}^{-k_2 - 3/2}
\]
for all sufficiently large $\n{x}$, and since $k_2 + 3/2 < 3$ this again
contradicts the integrability of $\rho_0$.
Thus only the alternative $E_0 < 0$ remains, and (\ref{compsupp}) 
is established.

Next we establish the desired regularity of the steady state.
Since $U_0 \in L^\infty (\R^3)$ this is also true for $\rho_0$,
cf.\ (\ref{rhodar}). This in turn implies that the first order
derivatives of $U_0$ are bounded, i.\ e.,
$U_0 \in W^{1,\infty}(\R^3)$. By Sobolev embedding
$U_0 \in C_b (\R^3)$, thus $\rho_0 \in C_c(\R^3)$. This in turn
implies that $U_0 \in C_b^1 (\R^3)$, thus $\rho_0 \in C_c^1(\R^3)$,
and thus $U_0 \in C_b^2 (\R^3)$. Observe that the function $h_\phi$
defined in (\ref{hdef}) is continuously differentiable.

If we define $V := - U_0 > 0$ and expand $1/\n{x-y}$ in powers of $y$
to third order for
$y \in \supp \rho_0$ and $\n{x}$ large we find that the
assumptions in \cite[Thm.~4]{GNN} hold. Thus $U_0$ is spherically symmetric
about some point in $\R^3$, and the proof of part (a) is complete.

As to part (b) we first observe that up to some shift $U_0$
as a function of the radial variable $r:=\n{x}$ solves the equation
\be \label{ef}
\frac{1}{r^2} (r^2 U_0')' = c_k (E_0 - U_0)_+^{k+3/2},\ r>0,
\ee
with some appropriately defined constant $c_k$. Here $'$ denotes the
derivative with respect to $r$.
The function $E_0 - U_0$ is a solution of the singular ordinary differential
equation 
\be \label{nef}
\frac{1}{r^2} (r^2 z')' = - c_k z_+^{k+3/2},\ r>0. 
\ee
Now observe that solutions
$z \in C ([0,\infty[) \cap C^2(]0,\infty[)$  of (\ref{nef}) with $z'$ bounded
near $r=0$ are uniquely determined by $z(0)$. This is due to the fact
that for such a solution the equation implies that $z'(0)$ exists and
is zero; clearly, $U_0'(0)=0$.   
Moreover, if $z$ is such a solution then so is
\[
z_{\alpha} (r) := \alpha z (\alpha^\gamma r),\ r \geq 0
\]
for any $\alpha >0$ where $\gamma := (k+1/2)/2$,  
and $z_\alpha (0) = \alpha z (0)$.
Now assume there exists another minimizer in $\F_M$, i.~e., up to a shift
another solution $U_1$ of (\ref{ef}) with cut-off energy $E_1 < 0$. 
Uniqueness for (\ref{nef}) yields some $\alpha >0$
such that 
\[
E_1 - U_1 (r) = \alpha E_0 - \alpha U_0(\alpha^\gamma r),\ r \geq 0 .
\]
However, both steady states have the same
total mass $M$, so that
\beas
M 
&=& 
c_k \int_0^\infty r^2 (E_1 - U_1 (r))_+^{k+3/2} dr \\
&=&
\alpha^{k+3/2 - 3 \gamma} c_k \int_0^\infty r^2 (E_0 - U_0 (r))_+^{k+3/2} dr
= \alpha^{k+3/2 - 3 \gamma} M.
\eeas
Since the exponent of $\alpha$ is not zero, this implies that $\alpha = 1$,
and considering limits at spatial infinity we conclude that $E_0=E_1$
and $U_0 = U_1$.
\prfe

\section{Dynamical stability}
\setcounter{equation}{0}

To investigate the dynamical stability of $f_0$ we note
that
\begin{equation}  \label{d-d}
\Hc (f)- \Hc (f_0)=d(f,f_0)-\frac{1}{8 \pi}
\|\nabla U_f-\nabla U_0\|^2_2,\ f \in \F_M
\end{equation}
where
\[
d(f,f_0) := \int\!\!\int \Bigl[Q(f)-Q(f_0) +
(E - E_0)(f-f_0)\Bigr]\,dv\,dx.
\]
Moreover, a simple Taylor expansion shows that 
\[ 
d(f,f_0) \geq 0,\ f \in \F_M
\]
always, and under further restrictions on $Q$, say for
$Q(f) = f^{1+1/k}$ with $1 \leq k < 3/2$, we even have
\[
d(f,f_0) \geq C \nn{f-f_0}_2^2,\ f \in \F_M .
\]

\begin{theorem} \label{stability}
Assume that the minimizer $f_0$ is unique in $\F_M$.
Then for every $\epsilon>0$ there is a $\delta>0$ such that
for any solution $t \mapsto f(t)$ of the Vlasov-Poisson system
with $f(0) \in C^1_c (\R^6)\cap \F_M$,
\[
d(f(0),f_0) + \frac{1}{8\pi} \|\nabla U_{f(0)}-\nabla U_0\|_2^2 < \delta
\]
implies that for every $t \geq 0$ there exists $a \in \R^3$ such that
\[
d(f^a(t),f_0) + \frac{1}{8\pi} \|\nabla U_{f^a(t)}-\nabla U_0\|_2^2 < \epsilon.
\]
\end{theorem}

\noindent
{\bf Proof.}
For $f(0) \in C^1_c (\R^6)\cap \F_M$ 
there exists a unique classical solution to the 
corresponding initial value problem, $f(t) \in \F_M,\ t \geq 0$, and
$\Hc$ is constant along $f(t)$.
 
Now assume the assertion of the theorem were false. 
Then there exist $\epsilon>0$, $t_n>0$, and
$f_n(0) \in C^1_c (\R^6)\cap \F_M$ such that 
\[
d(f_n(0),f_0) + 
\frac{1}{8\pi} \|\nabla U_{f_n(0)}-\nabla U_0\|_2^2 \to 0,\ n \to \infty,
\]
but
\be \label{cont}
\inf_{a \in \R^3} \left( d(f^a_n(t_n),f_0) +
\frac{1}{8\pi} \|\nabla U_{f^a_n(t_n)}-\nabla U_0\|_2^2 \right) 
\ge \epsilon,\ n \in \N.
\ee
Since $\Hc$ is conserved,  (\ref{d-d}) implies that  
\[
\lim_{n\to\infty} 
\Hc (f_n(t_n))=\lim_{n\to\infty} 
\Hc (f_n(0))=h_M,
\]
i.~e., $(f_n(t_n)) \subset \F_M$
is a minimizing sequence of $\Hc$. By Theorem~\ref{exminim} ,
we deduce that---up to a 
subsequence---$\|\nabla U_{{f^{a_n}_n}(t_n)}-\nabla U_0\|^2_2\to 0$. 
Since $(f_n^{a_n} (t_n))$ is a minimizing sequence
as well, (\ref{d-d}) implies that 
$d(f^{a_n}_n(t_n),f_0)\to 0$, a contradiction to (\ref{cont}). \prfe

\noindent
{\bf Final remarks}
\begin{itemize}
\item[(a)] We do in general have no control over the
shift vectors $a$. One might think that taking only initial
data with 
\be \label{centmass}
\int\!\!\int v f(x,v)\,dv\,dx = \int\!\!\int x f(x,v)\,dv\,dx = 0
\ee
might avoid the necessity of the shifts, since this condition
propagates and eliminates the trivial
instability due to perturbations of the form
$f_0(x+t V, v+V)$ with $V\in \R^3$ fixed. However,
as pointed out in \cite{Wa}, it is conceivable that
an appropriate, small perturbation causes a small fraction of the
total mass distribution to move off in one direction
and the bulk of the distribution in the other direction in such a way that
(\ref{centmass}) holds, but one still has to shift the reference frame with the bulk of the distribution to save the stability estimate.  
\item[(b)]
If we restrict the set $\F_M$ to spherically symmetric functions
then clearly all shift vectors $a=0$, and we recover the
results in \cite{GR1,GR2} for the $L$-independent case.
\item[(c)]
The question whether steady states which depend on
angular momentum $L$ are stable against nonsymmetric perturbations
remains open, since it is then no longer true that 
$\Hc$ is conserved along nonsymmetric solutions. 
\item[(d)]
Another open problem is the uniqueness of the minimizers
if $Q$ is not of the polytropic form (\ref{qpoly}).
We have found no substitute for the scaling argument used to analyse 
solutions of the equation (\ref{nef}) in the general case.
However, should the minimizer not be unique (not even locally)
then one still obtains a stability result in the sense that
the whole set of minimizers is stable, cf.\ \cite{GR1}.
\end{itemize}

\noindent
{\bf Acknowledgements:} The author would like to thank
{\sc A.~Unterreiter}, Universit\"at Kaiserslautern, for pointing 
out the crucial reference
\cite{L} to him. He also thanks {\sc Y.~Guo}, Brown University,
for helpful discussions.

\end{document}